\documentclass[journal,onecolumn,letterpaper,twoside,11pt]{IEEEtran}

\usepackage{cite}

\ifCLASSINFOpdf
  \usepackage[pdftex]{graphicx}
\else
\fi
\usepackage[cmex10]{amsmath}
\usepackage{array}
\ifCLASSOPTIONcompsoc
 \usepackage[caption=false,font=normalsize,labelfont=sf,textfont=sf]{subfig}
\else
  \usepackage[caption=false,font=footnotesize]{subfig}
\fi
\usepackage{fixltx2e}
\usepackage{dblfloatfix}

\ifCLASSOPTIONcaptionsoff
  \usepackage[nomarkers]{endfloat}
 \let\MYoriglatexcaption\caption
 \renewcommand{\caption}[2][\relax]{\MYoriglatexcaption[#2]{#2}}
\fi

\usepackage{url}

\usepackage{amsthm}
\usepackage{mathrsfs}
\usepackage{color}
\usepackage{mathtools,amssymb, nccmath}
\usepackage{bbm}
\usepackage{enumitem}
\setitemize{labelindent=0.4em,labelsep=0.2cm,leftmargin=0.85cm,topsep=2pt, partopsep=2pt,itemsep=2pt,parsep=2pt}

\theoremstyle{plain}

\newtheorem{lemma}{Lemma}
\newtheorem{problem}{Problem}
\newtheorem{corollary}{Corollary}
\newtheorem{remark}{Remark}

\newcommand{\ind}[1]{\mathbbm{1}_{\{#1\}}}   
\newcommand{\Exp}{{\mathsf{E}}}

\newcommand{\Hyp}{{\mathsf{H}}}
\newcommand{\sfb}{{\mathsf{b}}}
\newcommand{\f}{{\mathsf{f}}}
\newcommand{\g}{{\mathsf{g}}}
\newcommand{\p}{{\mathsf{p}}}
\newcommand{\sfd}{{\mathsf{d}}}
\newcommand{\sfr}{{\mathsf{r}}}
\newcommand{\sfu}{{\mathsf{u}}}
\newcommand{\sfL}{{\mathsf{L}}}
\newcommand{\sfS}{{\mathsf{S}}}
\newcommand{\sfG}{{\mathsf{r}}}
\newcommand{\A}{\mathsf{A}}
\newcommand{\B}{\mathsf{B}}
\newcommand{\T}{\mathsf{T}}
\newcommand{\sgn}{\mathrm{sign}}
\newcommand{\hJ}{{\hat{\mathcal{J}}}}
\newcommand{\Real}{\mathbb{R}}

\begin{document}

\title{Training Neural Networks for Likelihood/Density Ratio Estimation}

\author{George~V.~Moustakides  
and Kalliopi~Basioti 

\thanks{Manuscript received~~; revised ~~.}
\thanks{This work was supported by the US National Science Foundation under Grant CIF\,1513373 through Rutgers University.}
\thanks{G.\,V. Moustakides is with the Electrical and Computer Engineering Department, University of Patras, Rion, Greece, E-mail: moustaki@upatras.gr, and with the Department of Computer Science, Rutgers University, Piscataway, NJ, USA. E-mail: gm463@rutgers.edu.
}
\thanks{K. Basioti is a graduate student with the Department of Computer Science, Rutgers University, Piscataway, NJ, USA. E-mail: kib21@scarletmail. rutgers.edu.}
\thanks{Copyright (c) 2017 IEEE. Personal use of this material is permitted.  However, permission to use this material for any other purposes must be obtained from the IEEE by sending a request to pubs-permissions@ieee.org.}
}

\maketitle

\begin{abstract}
Various problems in Engineering and Statistics require the computation of the likelihood ratio function of two probability densities. In classical approaches the two densities are assumed known or to belong to some known parametric family. In a data-driven version we replace this requirement with the availability of data sampled from the densities of interest. For most well known problems in Detection and Hypothesis testing we develop solutions by providing neural network based estimates of the likelihood ratio or its transformations. This task necessitates the definition of proper optimizations which can be used for the training of the network. The main purpose of this work is to offer a simple and unified methodology for defining such optimization problems with guarantees that the solution is indeed the desired function. Our results are extended to cover estimates for likelihood ratios of conditional densities and estimates for statistics encountered in local approaches.
\end{abstract}


\IEEEpeerreviewmaketitle

\section{Introduction}\label{sec:1}

\IEEEPARstart{T}{he} likelihood ratio of two probability densities is a function that appears in a variety of problems in Engineering and Statistics. Characteristic examples \cite{poor}, \cite{moulin} constitute Hypothesis testing, Signal detection, Sequential hypothesis testing, Sequential detection of changes, etc. Many of these problems also use the likelihood ratio under a transformed form with the most frequent example being the log-likelihood ratio. In all these problems the main assumption is that the corresponding probability densities are available under some functional form. What we aim in this work is to replace this requirement with \textit{the availability of data sampled from each of the densities of interest}.

As we mentioned, the computation of the likelihood ratio function relies on the knowledge of the probability densities which, for the majority of applications, is an unrealistic assumption. One can instead propose parametric families of densities and, with the help of available data, estimate the parameters and form the likelihood ratio function. However, with the advent of Data Science and Deep Learning there is a phenomenal increase in need for processing data coming from images, videos etc. For most of these cases it is very difficult to propose any meaningful parametric family of densities that could reliably describe their statistical behavior. Therefore, these techniques tend to be unsuitable for most of these datasets.

If parametric families cannot be employed one can always resort to nonparametric density estimation \cite{silverman} and then form the likelihood ratio. These approaches are purely data-driven but require two different approximations, namely one for each density. Additionally, most nonparametric density estimators involve critical parameters, as the ``bandwidth'', that trade magnitude accuracy versus support accuracy with the corresponding tuning process not being well defined.

If we are only interested in the likelihood ratio, or its transformation with some fixed function, we could ask ourselves whether it is possible to directly estimate it, without passing through the double estimation of the two individual densities. Furthermore, motivated by their wide popularity and success in approximating nonlinear functions \cite{cybenko}, it is only natural to be interested in developing a methodology based on \textit{neural networks}. This is exactly the goal of this article, namely, to develop likelihood ratio estimation methods based on neural networks.

Of course, there already exists a substantial literature addressing the problem of \textit{density ratio estimation} \cite{sugiyama1,sugiyama2,kanamori} (and references therein). We must also mention the usage of these methods in several applications as covariate shift corrections \cite{stojanov}, density ratio estimation with dimensionality reduction \cite{sugiyama}, change detection \cite{khan}, approximate likelihood estimation \cite{izbicki}, feature selection \cite{braga}, etc. The focus and main tool in all these publications is density ratio estimation and no effort is made to estimate any \textit{transformation} of this ratio or any other meaningful statistic that occurs in Detection and Hypothesis testing. With our present work we intend to address these missing parts by developing a \textit{unified} methodology which can be used to directly estimate these functions.

Our paper is organized as follows: Section\,\ref{sec:1} contains the Introduction, In Section\,\ref{sec:2} we describe how a basic optimization problem must be defined in order to accept as solution a desired function. Section\,\ref{sec:3} contains a number of data-driven versions of the optimizations of Section\,\ref{sec:2} that are related to important problems in Detection and Hypotheses testing theory. In Section\,\ref{sec:4} we apply our methodology to specific problems from Hypothesis testing, Classification and Sequential change detection and compare the different possibilities. Finally, in Section\,\ref{sec:5} we present our concluding remarks.

\section{A General Optimization Problem}\label{sec:2}
Let us begin our analysis by introducing a simple optimization problem which will serve as our basis for the final, data-driven version. Suppose that $X$ is a random vector with probability density $\f_0(X)$ and consider the following cost function
\begin{equation}
\mathcal{J}(\sfu)=\Exp_{0}\left[\phi\big(\sfu(X)\big)+\sfr(X)\psi\big(\sfu(X)\big)\right],
\label{eq:cost0}
\end{equation}
where $\Exp_{0}[\cdot]$ denotes expectation with respect to $\f_0(X)$; $\phi(z),\psi(z)$ are scalar functions of the scalar $z$ and $\sfu(X),\sfr(X)$ are scalar functions of the vector $X$ with $\sfr(X)$ taking values in a \textit{known} interval $\mathcal{I}_{\sfr}$. We are interested in the following optimization problem 
\begin{equation}
\min_{\sfu(X)}\mathcal{J}(\sfu)=\min_{\sfu(X)}\Exp_{0}\big[\phi\big(\sfu(X)\big)+\sfr(X)\psi\big(\sfu(X)\big)\big].
\label{eq:optim0}
\end{equation}
Our goal is to \textit{design} the two functions $\phi(z),\psi(z)$ so that the solution of \eqref{eq:optim0} is of the form which is described next.

\begin{problem}\label{problem:1}
If $\omega(\sfr)$ is a {\bf\em known} scalar function of the scalar $\sfr$, we would like to design the two functions $\phi(z),\psi(z)$
so that the global minimizer in \eqref{eq:optim0} is equal to $\sfu(X)=\omega\big(\sfr(X)\big)$.
\end{problem}

Problem\,\ref{problem:1} is not difficult and with the next lemma we introduce a necessary condition which constitutes the first step towards its solution.

\begin{lemma}\label{lem:1}
In order for the optimization problem in \eqref{eq:optim0} to accept as solution the function $\omega\big(\sfr(X)\big)$, it is necessary that $\phi(z),\psi(z)$ satisfy
\begin{equation}
\phi'\big(\omega(\sfr)\big)+\sfr\psi'\big(\omega(\sfr)\big)=0,
\label{eq:cost5}
\end{equation}
for all real $\sfr\in\mathcal{I}_{\sfr}$.
\end{lemma}

\begin{IEEEproof}
Because $\sfu(X)$ is a function of $X$ the desired minimization can be performed as follows
\begin{equation}
\min_{\sfu(X)}\mathcal{J}(\sfu)=\min_{\sfu(X)}\Exp_{0}\left[\phi\big(\sfu(X)\big)+\sfr(X)\psi\big(\sfu(X)\big)\right]
=\Exp_{0}\left[\min_{\sfu(X)}\left\{\phi\big(\sfu(X)\big)+\sfr(X)\psi\big(\sfu(X)\big)\right\}\right].
\label{eq:cost7}
\end{equation}
In other words, we can apply the minimization directly onto the function under the expectation and perform it for each \textit{fixed} $X$. By fixing $X$, from \eqref{eq:cost7} we see that the optimum $\sfu(X)$ will depend only on the value of the function $\sfr(X)$ suggesting that the optimum is in fact of the form $\sfu\big(\sfr(X)\big)$. Because of this observation we can drop the dependence of $\sfu(X),\sfr(X)$ on $X$ and solve, instead, the following problem for each \textit{fixed} $\sfr\in\mathcal{I}_{\sfr}$
\begin{equation}
\min_{\sfu}\left\{\phi(\sfu)+\sfr\psi(\sfu)\right\}.
\label{eq:cost8}
\end{equation}
We know that all \textit{critical points} of $\phi(\sfu)+\sfr\psi(\sfu)$ satisfy the equation\footnote{Although not explicitly stated as assumption, wherever needed, we assume that derivatives and gradients of functions exist and are of sufficient smoothness.}
\begin{equation}
\phi'\big(\sfu(\sfr)\big)+\sfr\psi'\big(\sfu(\sfr)\big)=0.
\label{eq:cost9}
\end{equation}
Since we would like $\sfu(\sfr)=\omega(\sfr)$ to be a critical point, it is clear that \eqref{eq:cost9} must be true with $\sfu(\sfr)$ replaced by $\omega(\sfr)$, namely Eq.~\eqref{eq:cost5}. This completes the proof.
\end{IEEEproof}

\begin{remark}\label{rem:1}
In Lemma\,\ref{lem:1} it is particularly appealing the fact that \eqref{eq:cost5} is {\bf\em independent} from the probability density $\f_0(X)$ and the function $\sfr(X)$ and it only depends on $\omega(\sfr)$.
\end{remark}

\begin{corollary}
If the transformation $\omega(\sfr)$ is strictly monotone then \eqref{eq:cost5} is equivalent to
\begin{equation}
\phi'(z)+\omega^{-1}(z)\psi'(z)=0,
\label{eq:cost10}
\end{equation}
for all $z\in\omega(\mathcal{I}_{\sfr})$, where $\omega^{-1}(z)$ denotes the inverse function of $\omega(\sfr)$.
\end{corollary}

\begin{IEEEproof}
Eq.~\eqref{eq:cost10} follows immediately after defining $z=\omega(\sfr)$, which implies $\sfr=\omega^{-1}(z)$.
\end{IEEEproof}
\vskip0.2cm
\noindent\textbf{Assumption:}~\textit{From now on we concentrate on strictly monotone transformations $\omega(\sfr)$ and, therefore, we rely on \eqref{eq:cost10}. Furthermore, without loss of generality, we assume that $\omega(\sfr)$ is strictly increasing}.
\vskip0.2cm
Guaranteeing that the only solution in \eqref{eq:cost9} is the desired $\omega(\sfr)$ is possible by imposing additional constraints on the pair $\phi(z),\psi(z)$. The next lemma provides the required conditions.

\begin{lemma}\label{lem:1-2}
Select a function $\rho(z)$ that satisfies $\rho(z)<0$ for all $z\in\omega(\mathcal{I}_{\sfr})$ and define $\phi(z),\psi(z)$ with
\begin{equation}
\psi'(z)=\rho(z),~\text{and}~\phi'(z)=-\omega^{-1}(z)\rho(z),
\label{eq:derivs}
\end{equation}
then $\phi(z)+\sfr\psi(z)$ has a unique minimum for $z\in\omega(\mathcal{I}_{\sfr})$ at $z=\omega(\sfr)$. 
Furthermore, if $\mathcal{I}_{\sfr}=(0,\infty)$, i.e.~$\sfr>0$ and $\rho(z)$ is selected to satisfy
\begin{equation}
\rho'(z)>0~~\text{and}~~\omega^{-1}(z)\rho'(z)+\big(\omega^{-1}(z)\big)'\rho(z)<0,
\label{eq:convex}
\end{equation}
for all $z\in\omega\big((0,\infty)\big)$ then the function $\phi(z)+\sfr\psi(z)$ is strictly convex in $z$.
\end{lemma}

\begin{IEEEproof}
Using \eqref{eq:derivs} we can write
$$
\phi'(z)+\sfr\psi'(z)=\big(r-\omega^{-1}(z)\big)\rho(z).
$$
The roots of the equation $\phi'(z)+\sfr\psi'(z)=0$ are the critical points of $\phi(z)+\sfr\psi(z)$.
Clearly $z=\omega(\sfr)$ is the root of interest. In order to guarantee that no other root exists we need $\rho(z)\neq0$ for all $z\in\omega(\mathcal{I}_{\sfr})$. For this to be true, if we assume that $\rho(z)$ is continuous, it is necessary and sufficient that $\rho(z)$ does not change sign. Because we assumed that $\omega(\sfr)$ is strictly increasing, the same monotonicity property is inherited by the inverse function $\omega^{-1}(z)$ therefore, in order for $z=\omega(\sfr)$ to be the only minimum of $\phi(z)+\sfr\psi(z)$, its derivative which is equal to $\big(r-\omega^{-1}(z)\big)\rho(z)$, must be negative for $z<\omega(\sfr)$ and positive for $z>\omega(\sfr)$. This is assured when $\rho(z)<0$. No other extrema are possible unless we allow $\rho(z)\to0$ as $z$ approaches the end points of $\omega(\mathcal{I}_{\sfr})$ and, in this case, these extrema will be (local) maxima.

If $\sfr>0$ then we can also force the function $\phi(z)+\sfr\psi(z)$ to be convex in $z$. For this property to hold we need both functions $\phi(z),\psi(z)$ to be convex, namely, $\phi''(z)>0,\psi''(z)>0$. These two inequalities, after taking into account \eqref{eq:derivs}, are equivalent to \eqref{eq:convex}. This completes the proof.
\end{IEEEproof}

\begin{figure}[!h]
\centerline{\includegraphics[width=8cm]{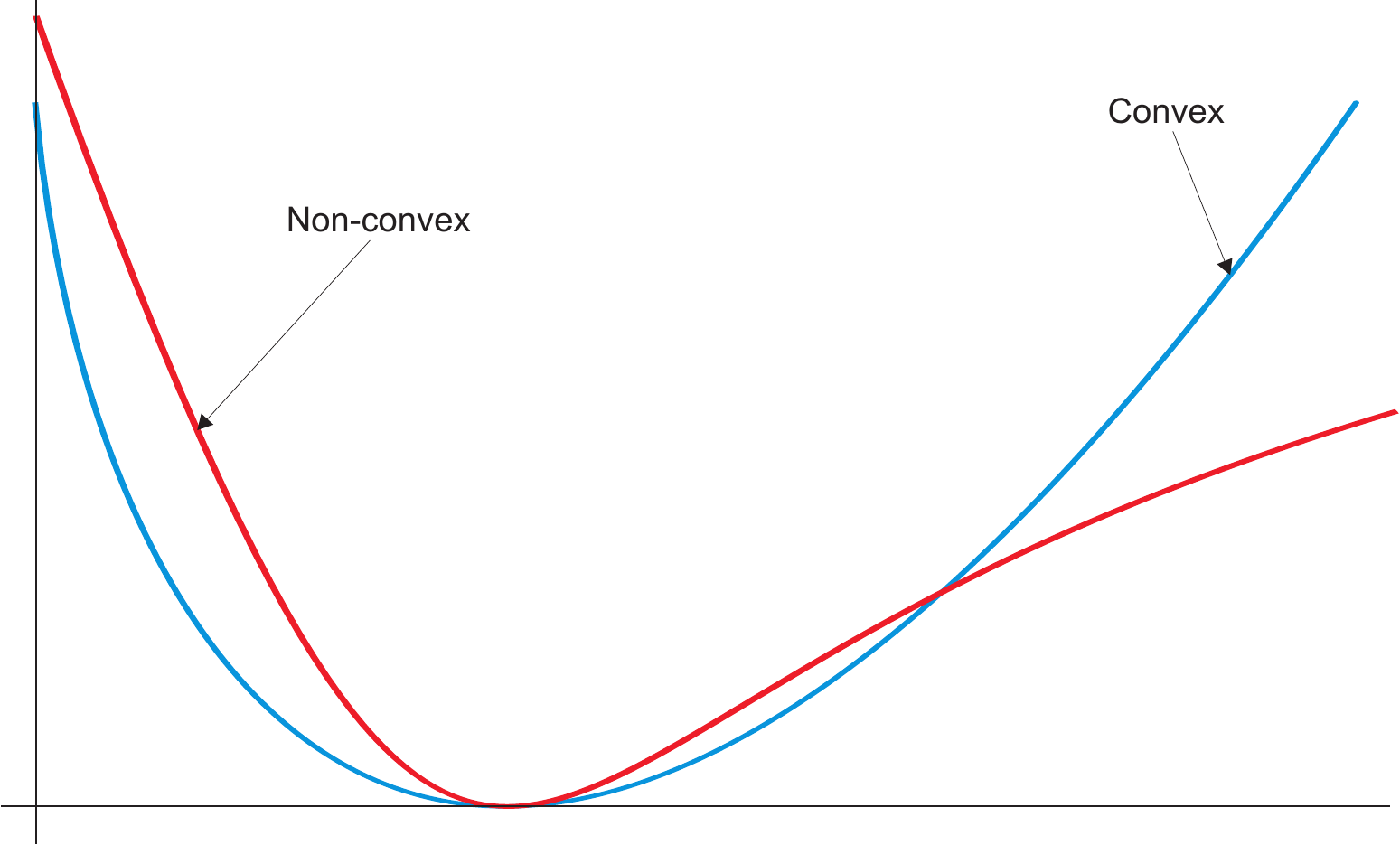}}
\vskip-0.2cm
\caption{Convex and non-convex forms of the function $\phi(z)+\sfr\psi(z)$ that exhibit a single minimum.}
\label{fig:1}
\end{figure}
Characteristic examples of convex (blue) and non-convex (red) functions enjoying a unique minimum are captured in Fig.\,\ref{fig:1}. The non-convex function can exhibit local maxima at the end points of its domain which in this example are 0 and $\infty$. Regarding the functions $\phi(z),\psi(z)$, they are defined in \eqref{eq:derivs}. This means that their explicit computation requires integration of $\rho(z)$ and $-\omega^{-1}(z)\rho(z)$ respectively. We have the following interesting point related to this issue.

\begin{remark}\label{rem:2}
For applying our method, as we explain in Section\,\ref{ssec:3.A}, we do not necessarily need closed form expressions for $\phi(z),\psi(z)$. This is because for the solution of the corresponding minimization problem we usually employ some form of gradient descent algorithm which requires only the derivatives and not the functions themselves. As we can see from \eqref{eq:derivs} these derivatives are defined in terms of $\rho(z),\omega(z)$ which are known.
\end{remark}

For different selections of the function $\omega(\sfr)$, let us present examples of pairs $\phi(z),\psi(z)$, derived by applying the previous analysis.

\subsection{Examples for $\omega(\sfr)=\sfr$}\label{ssec:2.A}
We are interested in identifying $\sfr(X)$ when no transformation is involved. This corresponds to selecting $\omega(\sfr)=\sfr$ and $\omega^{-1}(z)=z$. From our previous discussion, as long as $\rho(z)<0$, the optimum solution of \eqref{eq:cost8} is $\sfu(\sfr)=\sfr$ and the minimizer is unique. 

\noindent\underline{Case $\sfr(X)\in\Real$:} The function of interest can take any real value. We have $\mathcal{I}_{\sfr}=\Real$ and $\omega(\mathcal{I}_{\sfr})=\Real$, therefore we need to consider $z\in\Real$. According to \eqref{eq:derivs}, Lemma\,\ref{lem:1-2}, we must define
\begin{equation}
\phi'(z)=-z\rho(z),~~\psi'(z)=\rho(z).
\end{equation}
In all the examples that follow, one can verify that the corresponding function $\rho(z)$ is strictly negative (except at individual points).

\begin{itemize}
\item[A.1)] If we select $\rho(z)=-|z|^{\alpha}$, with $\alpha\neq-1,-2$ this yields $\phi(z)=\frac{|z|^{2+\alpha}}{2+\alpha}$ and $\psi(z)=-\frac{z|z|^{\alpha}}{1+\alpha}$. For $\alpha=-2$, $\rho(z)=-\frac{1}{z^2}$, $\phi(z)=\log|z|$, $\psi(z)=\frac{1}{z}$. For $\alpha=-1$, $\rho(z)=-\frac{1}{|z|}$, $\phi(z)=|z|$, $\psi(z)=-\mathrm{sign}(z)\log|z|$.

\item[A.2)] If we select $\rho(z)=-\frac{\tan^{-1}(z)}{z}$, this yields $\psi(z)=\int^z\rho(x)dx$ and $\phi(z)=z\tan^{-1}(z)-\frac{1}{2}\log(z^2+1)$.
This example corresponds to functions $\phi(z),\psi(z)$ that are not both available in closed form. However they can still be used to define an optimization problem whose solution is numerically computable.
\end{itemize}

\noindent\underline{Case $\sfr(X)>0$:} The function of interest is positive. This means $\mathcal{I}_{\sfr}=(0,\infty)$ and $\omega(\mathcal{I}_{\sfr})=(0,\infty)$, consequently we need to consider $z>0$. It is straightforward to see that all the previous examples are still valid with the simplification $|z|=z$ and $\mathrm{sign}(z)=1$. We have an additional example that applies to this case.

\begin{itemize}
\item[A.3)] If we select $\rho(z)=-\frac{1}{(1+z)z}$ then, $\phi(z)=\log(1+z)$ and $\psi(z)=\log(1+z^{-1})$.
\end{itemize}

Since $\sfr>0$ it is also possible to impose on $\rho(z)$ the constraints in \eqref{eq:convex} so that the resulting function $\phi(z)+\sfr\psi(z)$ is convex in $z$. These conditions take the form
\begin{equation}
\rho'(z)>0~~\text{and}~~\rho(z)+z\rho'(z)<0.
\label{eq:convex1}
\end{equation}
We can verify that in Example A.1 we enjoy convexity when $\alpha\in[-1,0]$ and that, we also enjoy convexity with Examples A.2 and A.3.

As we mentioned in the Introduction, in the existing literature the focus is primarily on the estimation of the likelihood ratio. In \cite{sugiyama1,sugiyama2} a general methodology is developed for defining optimization problems based on the Bregman divergence \cite{bregman}. In fact, we can show that there is a one-to-one correspondence between our function $\rho(z)$ and the function $f(t)$ adopted in \cite{sugiyama1} for the definition of the Bregman divergence, making the two approaches equivalent. However this equivalence applies only to $\omega(\sfr)=\sfr$ with $\sfr>0$ which covers the case of the likelihood ratio. Our method, as we emphasize repeatedly, is \textit{far more general}. Not only we can consider any nonlinear transformation $\omega(\sfr)$ but, even in the case $\omega(\sfr)=\sfr$, we can have $\sfr
\in\Real$ which, as we will see, allows us to estimate statistics encountered in local approaches that \textit{are not likelihood ratios}. For the existing literature, we would like to add that, the most popular optimization problem for likelihood ratio estimation is the mean square error
\begin{equation}
\mathcal{J}(\sfu)=\Exp_{0}[\big(\sfu(X)-\sfr(X)\big)^2]
=2\Exp_{0}\left[\frac{\sfu^2(X)}{2}-\sfr(X)\sfu(X)\right]+C.
\label{eq:MSE}
\end{equation}
Since the term $C=\Exp_{0}[\sfr^2(X)]$ does not contain the unknown function $\sfu(X)$, it can be omitted from the cost. The cost in \eqref{eq:MSE} corresponds to $\sfr(X)>0$ and it is a special case of Example\,A.1 with $\alpha=0$.

\subsection{Examples for $\omega(\sfr)=\log \sfr$}\label{ssec:2.B}
In this case we focus on $\log\sfr(X)$ meaning that $\omega(\sfr)=\log \sfr$, therefore $\omega^{-1}(z)=e^z$. Also we must have $\sfr(X)>0$ namely $\mathcal{I}_{\sfr}=(0,\infty)$ suggesting that $\omega\big((0,\infty)\big)=\Real$ which means that $z\in\Real$. As before $\rho(z)$ must be strictly negative in order to have a single minimum and, according to \eqref{eq:derivs} in Lemma\,\ref{lem:1-2}, we must define
\begin{equation}
\psi'(z)=\rho(z),~~\phi'(z)=-e^z\rho(z).
\label{eq:derivs4}
\end{equation}
In the examples that follow, we can verify that the corresponding function $\rho(z)$ is indeed negative.
\begin{itemize}
\item[B.1)] If $\rho(z)=-e^{-\alpha z}$ with $\alpha\neq0,1$, this produces
$$
\phi(z)=\frac{1}{1-\alpha}\{e^{(1-\alpha)z}-1\},~~\psi(z)=\frac{1}{\alpha}\{e^{-\alpha z}-1\}.
$$
If $\alpha\to0$ then
$\rho(z)=-1$, $\phi(z)=e^z$, $\psi(z)=-z$. If $\alpha\to1$ then $\rho(z)=-e^{-z}$, $\phi(z)=z$ and $\psi(z)=e^{-z}$. 

\item[B.2)] If $\rho(z)=-\frac{1}{1+e^z}$ then, $\phi(z)=\log(1+e^z)$ and $\psi(z)=\log(1+e^{-z})$.

\item[B.3)] If $\rho(z)=\frac{e^{-z}-1}{z}$ this yields $\phi(z)=\int^z\frac{e^x-1}{x}dx$ and $\psi(z)=\int^z\frac{e^{-x}-1}{x}dx$. The two functions $\phi(z),\psi(z)$ can be written in terms of the Exponential integral or with the help of a power series expansion, but they do not enjoy closed form expressions. On the other hand, their derivatives are simple and can be clearly used in an iterative algorithm to numerically compute the solution of the corresponding optimization.
\end{itemize}

Since $\sfr>0$ we can also consider the case of $\phi(z)+\sfr\psi(z)$ being convex. According to \eqref{eq:convex}, $\rho(z)$ is required, for $z\in\Real$, to satisfy
\begin{equation}
\rho'(z)>0~~\text{and}~~\rho(z)+\rho'(z)<0.
\label{eq:convex2}
\end{equation}
In Example B.1, \eqref{eq:convex2} is valid for $\alpha\in[0,1]$. In Examples B.2 and B.3 the particular selection of $\rho(z)$ also satisfies \eqref{eq:convex2}, thus giving rise to convex functions. There is no equivalent in the literature for this estimation problem and the corresponding estimation method appears here for the first time.

\subsection{Examples for $\omega(\sfr)=\frac{\sfr}{\sfr+1}$}\label{ssec:2.C}
For $\sfr(X)>0$ we are interested in the transformation $\omega(\sfr)=\frac{\sfr}{\sfr+1}$. If $\sfr(X)$ denotes likelihood ratio then $\omega\big(\sfr(X)\big)$ is the \textit{posterior probability} (for equal priors) which we can estimate directly. Again we have $\mathcal{I}_{\sfr}=(0,\infty)$ and therefore we must consider $z\in\omega\big((0,\infty)\big)=(0,1)$ while $\omega^{-1}(z)=\frac{z}{1-z}$. For the functions $\phi(z),\psi(z)$, from \eqref{eq:derivs} we have
\begin{equation}
\psi'(z)=\rho(z),~~~\phi'(z)=-\frac{z}{1-z}\rho(z).
\label{eq:aman2}
\end{equation}

\begin{itemize}
\item[C.1)] If we select $\rho(z)=-\frac{1}{z}$, this yields $\phi(z)=-\log(1-z)$ and $\psi(z)=-\log z$. This particular combination leads to the \textit{cross-entropy} loss method \cite{goodfellow} which, with the hinge loss method (presented in Section\,\ref{ssec:2.D}) constitute the most popular techniques for solving the classification problem. However, we must point out that from our analysis it becomes clear that the cross-entropy can also be used in other problems since it provides estimates of a strictly monotone transformation of the likelihood ratio.

\item[C.2)] If we select $\rho(z)=-(1-z)^\alpha$, for $\alpha\neq0,-1$, this yields $\phi(z)=-\frac{1}{\alpha(1+\alpha)}(az+1)(1-z)^{\alpha}$ and $\psi(z)=\frac{1}{1+\alpha}(1-z)^{1+\alpha}$. For $\alpha=0$ we obtain $\phi(z)=-z-\log(1-z)$ and $\psi(z)=-z$ and for $\alpha=-1$ the two functions become $\phi(z)=\log(1-z)+\frac{1}{1-z}$ and $\psi(z)=\log(1-z)$.
\end{itemize}

\subsection{Examples for $\omega(\sfr)=\mathrm{sign}(\log\sfr)$}\label{ssec:2.D} 
Here we are interested in estimating the sign of $\log\sfr$. This function appears in binary classification. Unfortunately $\mathrm{sign}(\log \sfr)$ is not continuous and not strictly increasing. The means to bypass this problem is to approximate the sign with a function that enjoys these necessary properties. We propose two different approximations that result in drastically different choices for $\phi(z),\psi(z)$.

\subsubsection*{\underline{Monotone Loss}} As a first approximation we propose $\mathrm{sign}(z)\approx\tanh(\frac{c}{2}z)$ where $c>0$ a parameter. We note that $\lim_{c\to\infty}\tanh(\frac{c}{2}z)=\mathrm{sign}(z)$. Using this approximation we can write for $\sfr>0$
\begin{equation}
\mathrm{sign}(\log \sfr)\approx\tanh\left(\frac{c}{2}\log \sfr\right)=\frac{\sfr^c-1}{\sfr^{c}+1}=\omega(\sfr).
\label{eq:sign}
\end{equation}
As we mentioned, we have exact equality for $c\to\infty$. Let us perform our analysis by assuming that $c$ is bounded. We note that $\omega^{-1}(z)=(\frac{1+z}{1-z})^{\frac{1}{c}}$. Since $\sfr>0$ this means that $\mathcal{I}_{\sfr}=(0,\infty)$ and $\omega(\mathcal{I}_{\sfr})=(-1,1)$ suggesting that we need to consider $z\in(-1,1)$. According to our general theory, in order to enjoy a unique solution in the minimization problem, we must select $\rho(z)<0$ for $z\in(-1,1)$ and, following \eqref{eq:derivs}, we must set
\begin{equation}
\psi'(z)=\rho(z),~~~\phi'(z)=-\left(\frac{1+z}{1-z}\right)^{\frac{1}{c}}\rho(z).
\label{eq:aman}
\end{equation}

\begin{itemize}
\item[D.1)] If we select $\rho(z)=-(1-z)^{\frac{1}{c}}$, this yields $\phi(z)=\frac{c}{1+c}(1+z)^{(\frac{1}{c}+1)}$ and $\psi(z)=\frac{c}{1+c}(1-z)^{(\frac{1}{c}+1)}$. If now we let $c\to\infty$ we obtain $\phi(z)=1+z$ and $\psi(z)=1-z$. Actually, we can discard the two constants since they do not affect the minimization problem and we can use instead $\phi(z)=-\psi(z)=z$ which results in \textit{linear loss}.

\item[D.2)] We can generalize the previous example. If we consider any $\rho(z)<0$ and define $\phi'(z),\psi'(z)$ according to \eqref{eq:aman}, then let $c\to\infty$, this will yield $\psi(z)=-\phi(z)$ with $\phi(z)$ being strictly increasing in $(-1,1)$ since $\phi'(z)=-\rho(z)>0$. This class of optimization problems was first proposed in \cite{basioti} by our group as a means to solve the binary classification problem. However, here it is derived as a special case for estimating the function $\sgn(\log\sfr)$.
\end{itemize}

\subsubsection*{\underline{Hinge Loss}} As a second approximation we can use $\mathrm{sign}(z)\approx \mathrm{sign}(z)|z|^c,~~c>0$, which is clearly increasing and continuous and converges to $\sgn(z)$ as $c\to0$. This suggests that
\begin{equation}
\mathrm{sign}(\log\sfr)\approx \mathrm{sign}(\log\sfr)|\log\sfr|^c=\omega(\sfr),
\label{eq:sign2}
\end{equation}
and $\omega^{-1}(z)=e^{\sgn(z)|z|^{\frac{1}{c}}}$. We have $\mathcal{I}_{\sfr}=(0,\infty)$ but now $\omega(\mathcal{I}_{\sfr})=\Real$ instead of the interval $(-1,1)$ we had in the previous case. This means that the functions $\phi(z),\psi(z)$ must be defined on the whole real line $z\in\Real$. We will present only one example that leads to a very well known optimization problem.

\begin{itemize}
\item[D.3)] Following \eqref{eq:derivs}, if we select $\psi'(z)=\rho(z)=-\{e^{-|z|^{\frac{1}{c}}}+ \ind{z<-1}\}<0$ then $\phi'(z)=e^{\sgn(z)|z|^{\frac{1}{c}}}\{e^{-|z| ^{\frac{1}{c}}}+\ind{z<-1}\}$. If we now let $c\to0$, we obtain the limiting form for the derivatives which become $\psi'(z)=-\ind{z<1}$ and $\phi'(z)=\ind{z>-1}$. By integrating we arrive at $\phi(z)=\max\{1+z,0\}$ and $\psi(z)=\max\{1-z,0\}$.
The optimization based on this particular pair is called the \textit{hinge loss} method \cite{tang}. The hinge loss and the cross-entropy loss methods, as we mentioned, are very popular in the classification problem where they are known to have excellent performance \cite{rosasco}, \cite{janocha}. 
\end{itemize}

\section{Data-driven Estimation based on Neural Networks}\label{sec:3}
In the previous section we introduced Problem\,\ref{problem:1} and described how to design the corresponding optimization problem so that it enjoys a predefined solution. In this section we apply this idea under a pure data-driven scenario. We target estimates for the likelihood and log-likelihood ratio function, but also for other known functions encountered in Detection, Hypothesis testing and Classification. 

To accommodate a data-driven setup we intend to apply two simultaneous approximations. The first consists in replacing statistical expectations with averages over available data and the second in replacing the general function $\sfu(X)$ with the output $\sfu(X,\theta)$ of a neural network\footnote{Although in our presentation we mention only neural networks, exactly the same analysis applies to any other function approximation method as for example kernel based. We can even accommodate the classical parametric density method. For example in the case of Gaussian modeling we can directly estimate the parameters of the likelihood ratio instead of the means and covariance matrices of the two densities.} with $\theta$ summarizing the network parameters. The first approximation is supported by the Law of Large Numbers whose validity, as we recall, \textit{does not} require independence of the data samples. For the second approximation, if the neural network is sufficiently rich then \cite{cybenko}, it can closely approximate any nonlinear function. Under these two approximations, the optimization problem in \eqref{eq:optim0} is transformed into a classical optimization over the parameters~$\theta$.

Even if we adopt the double approximation we just mentioned, due to the presence of $\sfr(X)$ which is unknown, it is still unclear how the cost function can be evaluated when we only have data. In what follows we introduce characteristic problems and explain how this difficulty can be resolved.

\subsection{Estimation of the Likelihood Ratio and its Transformations}\label{ssec:3.A}
In the cost $\mathcal{J}(\sfu)$ defined in \eqref{eq:cost0}, let $\sfr(X)$ denote the likelihood ratio $\sfr(X)=\frac{\f_1(X)}{\f_0(X)}$ of two probability densities $\f_0(X)$ and $\f_1(X)$.
Under this assumption we can rewrite the cost function as
\begin{equation}
\mathcal{J}(\sfu)=\Exp_{0}[\phi\big(\sfu(X)\big)+\sfr(X)\psi\big(\sfu(X)\big)]
=\Exp_{0}[\phi\big(\sfu(X)\big)]+\Exp_{1}[\psi\big(\sfu(X)\big)],
\label{eq:prob2}
\end{equation}
where $\Exp_{0}[\cdot],\Exp_{1}[\cdot]$ denote expectation with respect to the densities $\f_0(X),\f_1(X)$ respectively. Based on \eqref{eq:prob2} we can propose the following problem of interest.
\begin{problem}\label{problem:2}
We are given two sets of data samples $\{X^{0}_1,\ldots,X^{0}_{n_{0}}\}$ and $\{X^{1}_1,\ldots,X^{1}_{n_{1}}\}$ with the first set sampled from $\f_0(X)$ and the second from $\f_1(X)$. The two densities are {\bf\em unknown} and we are interested in finding a neural network based estimate of the transformation $\omega(\frac{\f_1(X)}{\f_0(X)})$ of the likelihood ratio function, where $\omega(\sfr)$ is a {\bf\em known} scalar function of the scalar $\sfr$.
\end{problem}

Following our methodology of Section\,\ref{sec:2} we first select the functions $\phi(z),\psi(z)$ which are appropriate for the desired transformation $\omega(\sfr)$, then we apply the two approximations mentioned in the beginning of this section, this yields
\begin{equation}
\mathcal{J}(\sfu)\approx\hJ(\theta)=
\frac{1}{n_{0}}\sum_{i=1}^{n_{0}}\phi\big(\sfu(X^{0}_i,\theta)\big)+
\frac{1}{n_{1}}\sum_{i=1}^{n_{1}}\psi\big(\sfu(X^{1}_i,\theta)\big).
\label{eq:newcost}
\end{equation}
Clearly, each data average approximates the corresponding expectation. Also with $\sfu(X,\theta)$ we limit $\sfu(X)$ to a class which can be represented by the output of a neural network. 
The optimization problem over the function $\sfu(X)$ defined in \eqref{eq:optim0} is now transformed into a classical optimization over the parameters $\theta$, that is,
\begin{equation}
\min_{\sfu(X)}\mathcal{J}(\sfu)\approx\min_{\theta}\hJ(\theta)=
\min_{\theta}\left\{\frac{1}{n_{0}}\sum_{i=1}^{n_{0}}\phi\big(\sfu(X^{0}_i,\theta)\big)+
\frac{1}{n_{1}}\sum_{i=1}^{n_{1}}\psi\big(\sfu(X^{1}_i,\theta)\big).
\right\}.
\label{eq:newoptim}
\end{equation}
When solved, \eqref{eq:newoptim} will produce an optimum $\theta_o$ which corresponds to a neural network that approximates the desired function, namely, $\sfu(X,\theta_o)\approx\omega\big(\sfr(X)\big)$ (provided the optimization algorithm does not converge to a local minimum, please also see the next remark). We realize that the new cost function $\hJ(\theta)$ uses only the two datasets and the pair of functions $\phi(z),\psi(z)$ which, as we pointed out, depend only on $\omega(\sfr)$ and require no knowledge of $\f_0(X),\f_1(X)$. Regarding \eqref{eq:newoptim} and its solution we have the following remark.

\begin{remark}\label{rem:4}
The local extrema in the general optimization problem become potential local extrema for the data-driven version, it is therefore imperative to remove them. As we have seen, this is achieved by assuring that the function $\phi(z)+\sfr\psi(z)$ has a single minimum for every $\sfr\in\mathcal{I}_{\sfr}$. We must, however, emphasize that this very desirable characteristic established for the general optimization problem is not necessarily carried over to the data-driven version in \eqref{eq:newoptim}. This is due to the structure of the neural network which gives rise to nonlinear costs that can still exhibit local minima.
\end{remark}

The optimization problem in \eqref{eq:newoptim} can be used to train the corresponding neural network. A possibility is to apply the gradient descent
\begin{equation}
\theta_t=\theta_{t-1}-\mu\left\{\frac{1}{n_{0}}\sum_{i=1}^{n_{0}}\phi'\big(\sfu(X^{0}_i,\theta_{t-1})\big)\nabla\!_\theta\sfu(X^{0}_i,\theta_{t-1})
+\frac{1}{n_{1}}\sum_{i=1}^{n_{1}}\psi'\big(\sfu(X^{1}_i,\theta_{t-1})\big)\nabla\!_\theta\sfu(X^{1}_i,\theta_{t-1})\right\},
\label{eq:alg1}
\end{equation}
where $\nabla\!_\theta$ denotes gradient with respect to $\theta$ and, as we can see, in each iteration we use all available data. Alternatively, when $n_0=n_1$ we can employ the stochastic gradient
\begin{equation}
\theta_t=\theta_{t-1}-\mu\left\{\phi'\big(\sfu(X^{0}_t,\theta_{t-1})\big)\nabla\!_\theta\sfu(X^{0}_t,\theta_{t-1})
+\psi'\big(\sfu(X^{1}_t,\theta_{t-1})\big)\nabla\!_\theta\sfu(X^{1}_t,\theta_{t-1})\right\}
\label{eq:alg2}
\end{equation}
where in each iteration $t$ we use a pair of samples $\{X^0_t,X^{1}_t\}$. In \eqref{eq:alg2}, when the data are exhausted we recycle them until the algorithm converges. In both updates $\mu>0$ denotes the step-size (learning rate).

From \eqref{eq:alg1}, \eqref{eq:alg2} we observe that we only need the derivatives $\phi'(z),\psi'(z)$ to perform the required training. Consequently, as pointed out in Remark\,\ref{rem:2}, the functions $\phi(z),\psi(z)$ are not needed explicitly for the algorithmic implementation of our method. Regarding the function $\sfu(X,\theta)$ we must assure that the output of the neural network captures the interval $\omega\big(\mathcal{I}_{\sfr}\big)$ precisely. For example, when we estimate the likelihood ratio function then the output must be positive, suggesting that we need to apply a proper nonlinearity (the ReLU or ELU) in the last layer of the network. When, on the other hand, we estimate the log-likelihood ratio, which can be any real number, no such nonlinearity is necessary.

\subsubsection*{\underline{Estimation of the Kullback-Leibler and the Mutual Information Number}}
A side product of our method for estimating log-likelihood ratio function is the estimation of the Kullback-Leibler (K-L) information number and the mutual information number. The K-L information number between two densities $\f_0(X),\f_1(X)$ is defined as $I(\f_1,\f_0)=\Exp_{1}[\log\frac{\f_1(X)}{\f_0(X)}]$. If $\sfu(X,\theta_o)$ is the resulting neural network obtained from solving \eqref{eq:newoptim} for the estimation of $\log\frac{\f_1(X)}{\f_0(X)}$, then
\begin{equation}
I(\f_1,\f_0)\approx\hat{I}(\f_1,\f_0)=\frac{1}{n_{1}}\sum_{i=1}^{n_1}\sfu(X^{1}_i,\theta_o).
\label{eq:K-L}
\end{equation}
constitutes a straightforward estimate of the K-L information number.

For the mutual information, suppose we have two random vectors $X,Y$ that follow the joint density $\f(X,Y)$, then the mutual information is defined as
\begin{equation}
I(X;Y)=\Exp\left[\log\frac{\f(X,Y)}{\f_X(X)\f_Y(Y)}\right],
\label{eq:MI}
\end{equation}
where $\f_X(X),\f_Y(Y)$ denote the marginal densities of $X$ and $Y$ respectively.
To estimate $I(X;Y)$ we first estimate the log-likelihood ratio $\log\frac{\f(X,Y)}{\f_X(X)\f_Y(Y)}$ between the density $\f_0(X,Y)=\f_X(X)\f_Y(Y)$ which treats $X$ and $Y$ as statistically independent and $\f_1(X,Y)=\f(X,Y)$. This leads to the definition of the next problem.

\begin{problem}
We are given a {\bf\em single} set of data pairs $\{(X_1,Y_1),\ldots,(X_n,Y_n)\}$ that follow the {\bf\em unknown} joint density $\f(X,Y)$. We would like to use these data to obtain a neural network estimate of the log-likelihood ratio function $\log\frac{\f(X,Y)}{\f_X(X)\f_Y(Y)}$.
\end{problem}

To apply our method we must have two data sets. Clearly, the first is the existing which is sampled from $\f(X,Y)$ but we also need a second set sampled from $\f_X(X)\f_Y(Y)$. Since the $X$-components of the available pairs follow $\f_X(X)$ and the $Y$-components $\f_Y(Y)$, we can form a second set by simply combining \textit{each} $X_i$ with \textit{all} the $\{Y_j\}$ samples. This suggests the following form for \eqref{eq:newcost}
\begin{equation}
\hJ(\theta)=
\frac{1}{n^2}\sum_{j=1}^n\sum_{i=1}^{n}\phi\big(\sfu(X_i,Y_j,\theta)\big)+
\frac{1}{n}\sum_{i=1}^{n}\psi\big(\sfu(X_i,Y_i,\theta)\big),
\label{eq:newcost2}
\end{equation}
where the neural network $\sfu(X,Y,\theta)$ has as input both vectors $X,Y$. Once we optimize $\hJ(\theta)$ in \eqref{eq:newcost2} and $\sfu(X,Y,\theta_o)$ is the resulting neural network, we estimate the mutual information as
$$
I(X;Y)\approx\hat{I}(X;Y)=\frac{1}{n}\sum_{i=1}^n\sfu(X_i,Y_i,\theta_o).
$$

\subsection{Estimation of the Log-Likelihood Ratio of Conditional Densities}\label{ssec:3.B}
For a time series $\{x_t\}$ it is often required to compute the log-likelihood ratio of all the data samples $\{x_1,\ldots,x_t\}$ that are available up to time $t$. In on-line processing this computation must be repeated every time a new sample is acquired. We assume that the data can follow two hypotheses described by two probability densities $\f_0(\cdot),\f_1(\cdot)$. Since $\f_0(x_t,\ldots,x_1)=\f_0(x_t|x_{t-1},\ldots,x_1)\f_0(x_{t-1},\ldots,x_1)$, and similarly for $\f_1(\cdot)$, this results in the update
\begin{equation}
\sfL_t=\sfL_{t-1}+\log\frac{\f_1(x_t|x_{t-1},\ldots,x_1)}{\f_0(x_t|x_{t-1},\ldots,x_1)},
\label{eq:L1}
\end{equation}
where $\sfL_t=\log\frac{\f_1(x_t,\ldots,x_1)}{\f_0(x_t,\ldots,x_1)}$ is the log-likelihood ratio of the samples up to time $t$ and $\f_0(x_t|x_{t-1},\ldots,x_1)$, $\f_1(x_t|x_{t-1},\ldots,x_1)$ are the conditional densities of $x_t$ given the past samples $x_{t-1},\ldots,x_1$.

The Markovian is clearly one of the most important models for time series. If we assume that under both densities the process $\{x_t\}$ is \textit{homogeneous Markov} of order $k$ then for $t>k$ we can write
$$
\f_0(x_t|x_{t-1},\ldots,x_1)=\f_0(x_t|x_{t-1},\ldots,x_{t-k}),
$$
and a similar expression applies for $\f_1(\cdot)$, suggesting that the update in \eqref{eq:L1} for $t>k$, becomes
\begin{equation}
\sfL_t=\sfL_{t-1}+\log\frac{\f_1(x_t|x_{t-1},\ldots,x_{t-k})}{\f_0(x_t|x_{t-1},\ldots,x_{t-k})}.
\label{eq:L2}
\end{equation}
Using the definition of the conditional probability density we can also write
\begin{equation}
\log\frac{\f_1(x_t|x_{t-1},\ldots,x_{t-k})}{\f_0(x_t|x_{t-1},\ldots,x_{t-k})}=
\log\frac{\f_1(x_t,\ldots,x_{t-k})}{\f_0(x_t,\ldots,x_{t-k})}-\log\frac{\f_1(x_{t-1},\ldots,x_{t-k})}{\f_0(x_{t-1},\ldots,x_{t-k})}.
\label{eq:L3}
\end{equation}
The previous equality implies that we can apply the following approximation to the log-likelihood function of the conditional densities
\begin{equation}
\log\frac{\f_1(x_t|x_{t-1},\ldots,x_{t-k})}{\f_0(x_t|x_{t-1},\ldots,x_{t-k})}\approx
\sfu_{k+1}(x_t,\ldots,x_{t-k},\theta_{k+1})-\sfu_k(x_{t-1},\ldots,x_{t-k},\theta_k)
\label{eq:L4}
\end{equation}
where 
\begin{align}
\begin{split}
\sfu_k(x_{t},\ldots,x_{t-k+1},\theta_k)&\approx\log\frac{\f_1(x_{t},\ldots,x_{t-k+1})}{\f_0(x_{t},\ldots,x_{t-k+1})}\\
\sfu_{k+1}(x_t,\ldots,x_{t-k},\theta_{k+1})&\approx\log\frac{\f_1(x_t,\ldots,x_{t-k})}{\f_0(x_t,\ldots,x_{t-k})}
\end{split}
\label{eq:L11}
\end{align}
and $\sfu_j(x_j,\ldots,x_1,\theta_j)$ denotes the approximation of the $j$th order log-likelihood ratio with $\theta_j$ summarizing the coefficients of the corresponding neural network. 

If we are given a realization $\{x^{0}_1,\ldots,x^{0}_{n_{0}}\}$ of the process under $\f_0(\cdot)$ and a realization $\{x^{1}_1,\ldots,x^{1}_{n_{1}}\}$ under $\f_1(\cdot)$, then the two functions $\sfu_k(\cdot),\sfu_{k+1}(\cdot)$ in \eqref{eq:L11} can be obtained with the help of two separate optimization problems as described in Section\,\ref{ssec:3.A}. The data vectors $\{X^{0}_t\},\{X^{1}_t\}$ are defined as $X^{0}_t=[x^{0}_t\cdots x^{0}_{t-k}]^\intercal$, $X^{1}_t=[x^{1}_t\cdots x^{1}_{t-k}]^\intercal$ for the estimate of the $(k+1)$st order log-likelihood ratio and $X^{0}_t=[x^{0}_t\cdots x^{0}_{t-k+1}]^\intercal$, $X^{1}_t=[x^{1}_t\cdots x^{1}_{t-k+1}]^\intercal$ for the estimate of the $k$th order.

Once $\sfu_k(x_k,\ldots,x_1,\theta_k)$, $\sfu_{k+1}(x_{k+1},\ldots,x_1,\theta_{k+1})$ have been designed, following \eqref{eq:L2}, we can use them to update an estimate $\hat{\sfL}_t$ of $\sfL_t$ 
\begin{equation*}
\hat{\sfL}_t=\hat{\sfL}_{t-1}+
\sfu_{k+1}(x_t,\ldots,x_{t-k},\theta_{k+1})-\sfu_k(x_{t-1},\ldots,x_{t-k},\theta_k).
\end{equation*}
This idea can be very useful in hypothesis testing for the implementation of the sequential probability ratio test (SPRT) \cite{wald}. We recall that SPRT is used to sequentially decide between two hypotheses. With the previous update we can implement it without the need to know the corresponding densities. 

Estimates of the log-likelihood ratio of conditional probability densities also allow for the recursive computation of the cumulative sum (CUSUM) statistic \cite{page} which is used to sequentially detect a change in the statistical behavior of a process. If $\sfS_t$ denotes the CUSUM statistic and $\hat{\sfS}_t$ its estimate then for the two quantities we can write the following update for $t>k$
\begin{align}
\begin{split}
\sfS_t&=\sfS_{t-1}^++\log\frac{\f_1(x_t|x_{t-1},\ldots,x_{t-k})}{\f_0(x_t|x_{t-1},\ldots,x_{t-k})}\allowdisplaybreaks\\ \allowdisplaybreaks
\hat{\sfS}_t&=\hat{\sfS}_{t-1}^++
\sfu_{k+1}(x_t,\ldots,x_{t-k},\theta_{k+1})-\sfu_k(x_{t-1},\ldots,x_{t-k},\theta_k),
\end{split}
\label{eq:L12}
\end{align}
where $x^+=\max\{x,0\}$. If we are willing to sacrifice the first $k$ time instances then we can initialize with $\sfS_k=\hat{\sfS}_k=0$. We also recall that $\f_0(\cdot)$ corresponds to the pre- and $\f_1(\cdot)$ to the post-change probability density.

\subsection{Estimation of Local Statistics}\label{ssec:3.C}
In Detection theory there is interest in problems where the nominal probability density is given but the alternative is known only as a parametric density. In this case a possible solution for testing/detection can be the application of the \textit{local approach} (an alternative would be the GLRT, which we discuss in Section\,\ref{ssec:4.C}) and the usage of the \textit{locally most powerful} (LMP) test \cite[Pages 51--52]{poor},\cite[Pages 82--83]{moulin}. 

The local formulation, as we pointed out, relies on a parametric density $\f_1(X,\vartheta)$ where the nominal density corresponds (without loss of generality) to $\vartheta=0$, i.e.~
\begin{equation}
\f_0(X)=\f_1(X,0).
\label{eq:local00}
\end{equation}
If we are interested in distinguishing between $\vartheta=0$ and $\vartheta\neq0$ with the alternative $\vartheta$ being unknown, one can examine the case $\|\vartheta\|\ll1$, which corresponds to a ``difficult'' scenario due to the closeness of the two densities. The ``small'' size of $\vartheta$ allows for the analysis of the limiting form of the likelihood ratio, after Taylor expanding it
\begin{equation}
\frac{\f_1(X,\vartheta)}{\f_0(X)}\approx 1+\frac{\vartheta^\intercal \nabla\!_{\vartheta}\f_1(X,\vartheta)|_{\vartheta=0}}{\f_0(X)}.
\label{eq:local1}
\end{equation}
If we attempt to distinguish between $\vartheta=0$ and $\vartheta$ having a value $\vartheta=\epsilon\delta$ where $\delta$ is a known direction and $\epsilon>0$ is unknown but ``small'',
then for testing we can use the \textit{local statistic}
\begin{equation}
\sfr(X)=\frac{\delta^\intercal\nabla\!_{\vartheta}\f_1(X,\vartheta)|_{\vartheta=0}}{\f_0(X)},
\label{eq:local2}
\end{equation}
which must be compared against a constant threshold to make a decision. This corresponds to the LMP test \cite{moulin,poor}.

We would like to consider this problem under a data-driven setup. In order to free ourselves from the need of a specific parametric family, we focus on the case where \eqref{eq:local2} results in the following special form
\begin{equation}
\sfG(X)=\sfd(X)+[\p_1(X)\cdots\p_k(X)]\frac{\nabla\!_{X}\f_0(X)}{\f_0(X)},
\label{eq:GX}
\end{equation}
with $\sfd(X)$, $\p_l(X),l=1,\ldots,k$ are \textit{known} functions of $X$ and $k$ denotes the size of $X$.
Characteristic example where the local statistic satisfies \eqref{eq:GX} is $\f_1(X,\vartheta)=\f_0(X-\vartheta)$ where we translate the data by a vector $\vartheta$ and the local statistic for $\vartheta=\epsilon\delta$ becomes $-\delta^\intercal\frac{\nabla\!_{X}\f_0(X)}{\f_0(X)}$ with $\sfd(X)=0$ and $\p_l(X)=-\delta_l$ where $\delta=[\delta_1\cdots\delta_k]^\intercal$. A second example is when $\f_1(X,\vartheta)=\f_0\big((1+\vartheta)X\big)(1+\vartheta)$ with $\vartheta$ scalar, producing a change in scale, and the resulting local statistic becomes $1+X^\intercal\frac{\nabla\!_{X}\f_0(X)}{\f_0(X)}$ with $\sfd(X)=1$ and $\p_l(X)=x_l$ where $X=[x_1,\ldots,x_k]^\intercal$.  More generally, if we start with a random vector $X$ that follows $\f_0(X)$ and apply a transformation\footnote{In fact \textit{any} nonlinear transformation $T(X,\vartheta)$ that satisfies \eqref{eq:local00}, under proper smoothing conditions and for ``small'' $\vartheta$ becomes equivalent to this form provided that $\nabla\!_{\vartheta}T(X,\vartheta)|_{\vartheta=0}$ is not identically 0.} of the form $Y=X+A(X)\vartheta$, where $A(X)$ is a \textit{known} matrix function of $X$, the resulting random vector $Y$ will have a density which is parametrized in $\vartheta$. It is not hard to show that the local statistic for this density satisfies \eqref{eq:GX}. After these observations, we arrive at our last problem of interest.

\begin{problem}\label{problem:3}
We are given a {\bf\em single} set of data samples $\{X_1,\ldots,X_{n}\}$ which follow the {\bf\em unknown} probability density $\f_0(X)$. We would like, using the available dataset, to design a neural network estimate of the function $\sfG(X)$ defined in \eqref{eq:GX} where $\sfd(X),\p_l(X),l=1,\ldots,k$ are {\bf\em known} functions of~$X$.
\end{problem}

To apply the method developed in Section\,\ref{ssec:2.A} we need to find a means to compute the cost $\mathcal{J}(\sfu)$ for $\sfr(X)$ satisfying \eqref{eq:GX}. 
We first note that
\begin{multline}
\Exp_0\left[\frac{\frac{\partial\f_0(X)}{\partial x_l}}{\f_0(X)}\p_l(X)\psi(\sfu(X))\right]=\int \frac{\partial\f_0(X)}{\partial x_l}\p_l(X)\psi(\sfu(X))dX\\
= -\Exp_0\left[\frac{\partial\p_l(X)}{\partial x_l}\psi\big(\sfu(X)\big)+\p_l(X)\frac{\partial\sfu(X)}{\partial x_l}\psi'\big(\sfu(X)\big)\right],
\label{eq:mbarouak1}
\end{multline}
where for the last equality we used integration by parts and assumed that $\f_0(X)\p_l(X)\psi(\sfu(X))\to0$ when $x_l\to\pm\infty$ for all $l=1,\ldots,k$. This assumption is not unrealistic since well behaved densities $\f_0(X)$ satisfy $\f_0(X)\to0$ when $x_l\to\pm\infty$. Using \eqref{eq:mbarouak1} we can rewrite the cost function as
\begin{multline}
\mathcal{J}(\sfu)=\Exp_0[\phi\big(\sfu(X)\big)+\sfr(X)\psi\big(\sfu(X)\big)]\\
=\Exp_0\bigg[\phi\big(\sfu(X)\big)+\psi\big(\sfu(X)\big)
\left(\sfd(X)-\sum_{l=1}^k\frac{\partial\p_l(X)}{\partial x_i}\right)
-\psi'\big(\sfu(X)\big)[\p_1(X)\cdots\p_k(X)]\nabla\!_{X}\sfu(X)\bigg],
\label{eq:mbarouak3}
\end{multline}
and now, under the expectation we have only known terms. This means that we can approximate \eqref{eq:mbarouak3} using the available dataset $\{X_1,\ldots,X_n\}$. Indeed, we can write
\begin{equation}
\mathcal{J}(\sfu)\approx\hat{\mathcal{J}}(\theta)=\frac{1}{n}\sum_{i=1}^n{\mit\Omega}(X_i,\theta)
\label{eq:Omega}
\end{equation}
where
\begin{multline}
~~~~{\mit\Omega}(X,\theta)=
\phi\big(\sfu(X,\theta)\big)+\psi\big(\sfu(X,\theta)\big)\left(\sfd(X)-\sum_{i=1}^k\frac{\partial\p_i(X)}{\partial x_i}\right)\\
-\psi'\big(\sfu(X,\theta)\big)[\p_1(X)\cdots\p_k(X)]\nabla\!_{X}\sfu(X,\theta).~~~~
\label{eq:mbarouak4}
\end{multline}
Since $\sfr(X)\in\Real$, we need to select $\phi(z),\psi(z)$ from Examples\,A.1 or A.2 that are developed for this case. As before, for the minimization of the corresponding cost we can apply gradient descent
\begin{equation}
\theta_t=\theta_{t-1}-\mu\sum_{l=1}^n\nabla\!_\theta{\mit\Omega}(X_l,\theta_{t-1}),
\label{eq:alg1-1}
\end{equation}
or stochastic gradient
\begin{equation}
\theta_t=\theta_{t-1}-\mu\nabla\!_\theta{\mit\Omega}(X_t,\theta_{t-1}).
\label{eq:alg2-2}
\end{equation}
This completes the presentation of all our problems of interest.


\section{Experiments}\label{sec:4}
In Section\,\ref{sec:2} we introduced examples of optimization problems that accept as solutions transformations of the likelihood ratio and in Section\,\ref{sec:3} we presented how these problems can be implemented under a data-driven case. In this section we will apply our methodology to a number of problems related to Detection and Hypothesis testing. Basic element in our simulation constitutes the neural network which, in all our experiments, will be a simple full 2-layer network (1 hidden layer) with a single output. Of course one can use more layers if necessary and networks of special form, as convolutional. The equations that define the network structure are the following
\begin{equation}
Z=\A_1X+\A_0,~~U=\g_1(Z),~~v=\B_1^\intercal U+\sfb_0,~~\sfu=\g_0(v),
\label{eq:nn}
\end{equation}
where $X$ is the input vector of length $k$; $\A_1$ is a matrix of size $N\times k$ with $N$ the length of the hidden layer; $\A_0,Z,U$ are vectors of length $N$ with $\A_0$ containing the offsets of the hidden layer; $\g_1(z)$ is the activation function which is a scalar function of the scalar $z$ with $\g_1(Z)$ meaning that $\g_1(z)$ is applied to each element of the vector $Z$; $\B_1$ is a vector of length $N$ that combines the hidden layer elements; $v,\sfb_0,\sfu$ are scalars with $\sfb_0$ being the offset of the last layer and $\g_0(v)$ is a scalar function that forces the network output $\sfu$ to lie in the appropriate interval. It is clear that the parameter vector $\theta$ mentioned in our analysis is the collection $\theta=\{\A_0,\A_1,\sfb_0,\B_1\}$ which involves $N(k+2)+1$ parameters that need to be identified through training (by solving the corresponding optimization problem).

In all our examples, for the hidden layer activation function we select the ReLU, namely, $\g_1(z)=\max\{z,0\}$, which is the most popular in applications. The configuration of the neural network is denoted as $k\times N\times 1$ where, as we mentioned, $k$ is the input size, $N$ the size of the hidden layer and 1 is for the output size. Let us now give specific details about the methods we intend to test.

\subsubsection*{\underline{Mean Square Loss for Likelihood Ratio Estimation}} Following Section\,\ref{ssec:2.A} we select $\phi(z)=\frac{z^2}{2}$ and $\psi(z)=-z$, that correspond to Example\,A.1 with $\alpha=0$ (and $\sfr>0$). This is the most popular method in the literature for likelihood ratio estimation. We also need to select $\g_0(z)$ in order to make the output non-negative. This turns out to be a difficult issue. We can select $\g_0(z)$ to be the ReLU, namely set all negative $v$ values in \eqref{eq:nn} to 0. Unfortunately, the resulting network tends to generate negative $v$ values frequently and when they are set to 0, this results in a peculiar behavior when, for example, we use it in hypothesis testing.

To remedy this problem, instead of setting the negative values to 0 it is preferable to replace them with some exponentially small positive values by using the function $\g_0(z)=c e^{z}\ind{z\leq0}+(z+c)\ind{z\geq0}$ where $c>0$ some small constant. This nonlinearity is known as the ELU. In any case, it is clear that the non-negativity constraint requires special attention. We must also mention that similar problem is observed when, instead of neural networks, we adopt other forms of approximation as kernel based \cite[Pages 68--70]{sugiyama2} where, again, incorporating the non-negativity property in the optimization is not a simple task. Interestingly, this crucial constraint is satisfied \textit{for free} if instead of the likelihood ratio we estimate the log-likelihood ratio, as we propose next.

\subsubsection*{\underline{Exponential Loss for Log-Likelihood Ratio Estimation}} Following our methodology in Section\,\ref{ssec:2.B}, Example\,B.1, for the log-likelihood ratio estimation, we use of $\phi(z)=e^{0.5z}$, $\psi(z)=e^{-0.5z}$, corresponding to $\alpha=0.5$. In this case there is no need for any output nonlinearity $\g_0(z)$ since the log-likelihood ratio takes values in $\Real$. In fact exactly this property makes the estimate of the log-likelihood ratio attractive for also estimating the likelihood ratio. Indeed, if $\sfu(X)$ denotes the estimate of the log-likelihood ratio then $e^{\sfu(X)}$ can play the role of the estimate of the likelihood ratio. As we can see, \textit{we satisfy the non-negativity constraint with absolutely no effort}. 

\subsubsection*{\underline{Cross-Entropy Loss for Posterior Probability Estimation}} From Section\,\ref{ssec:2.C}, Example\,C.1, we select the cross-entropy method with $\phi(z)=-\log(1-z)$, $\psi(z)=-\log(z)$. Here the output must lie in the interval $(0,1)$ for this reason we select the sigmoid $\g_0(z)=\frac{e^z}{1+e^z}$ as the output nonlinearity. Although cross-entropy is known for classification it can be clearly used in any other problem which requires the likelihood or log-likelihood ratio.

\subsubsection*{\underline{Linear and Hinge Loss for Sign Log-Likelihood Estimation}} We first refer to Section\,\ref{ssec:2.D}, Example D.1. We select $\phi(z)=-\psi(z)=z$ but we must limit the output of the network within the interval $[-1,1]$. We can do this either by using the function $\g_0(z)=\tanh(z)$ or by employing a nonlinear transformation that \textit{is not necessarily one-to-one}. Following \cite{basioti}, possibilities are $\g_0(z)=\frac{s z}{s-1+|z|^s},s>1$ or $\g_0(z)=ze^{\frac{1}{s}(1-|z|^s)},s>0$. Both functions are almost linear for $z\in[-1,1]$ while the values $|z|>1$ are again mapped back to $[-1,1]$. For our simulation we use the first case with $s=2$ resulting in $\g_0(z)=\frac{2z}{1+z^2}$.

For classification, especially in the binary case, there are numerous possibilities. Particular attention has attracted the technique based on the hinge loss \cite{tang} which we also presented in Section\,\ref{ssec:2.D}, Example\,D.3, and known to enjoy an overall very satisfactory performance \cite{rosasco}, \cite{janocha}. Here we have $\phi(z)=\max\{1+z,0\}$, $\psi(z)=\max\{1-z,0\}$ with $z\in\Real$ consequently, unlike our previous method, no nonlinearity $\g_0(z)$ is needed in the output.

\begin{remark} Specific combinations of the functions $\phi(z),\psi(z)$ and $\g_0(z)$ can establish equivalence between different approximation problems. For instance, the estimation of the log-likelihood ratio with $\phi(z),\psi(z)$ from Example\,B.2 and no $\g_0(z)$; the estimation of the likelihood ratio with $\phi(z),\psi(z)$ from Example\,A.3 with $\g_0(z) = e^z$ and the estimation of the posterior probability with $\phi(z),\psi(z)$ from Example\,C.1 (cross-entropy) with $\g_0(z)=\frac{e^z}{1+e^z}$, are equivalent in the sense that if the network parameters are initialized in the same way, at each iteration of the training algorithm, the updates of the network parameters are exactly the same.
\end{remark}

We are now ready to apply all these different methods to a number of known problems from Detection and Hypothesis testing. We will mostly use synthetic data in order to be able to compare with the optimum schemes but we also include an interesting example that uses the MNIST database with images of handwritten numerals.

\subsection{Application to Binary Hypothesis Testing}\label{ssec:4.A}
Suppose $X$ is a random vector $X$ and we have the following two hypotheses
\vskip0.1cm
\noindent
\hskip0.5cm$\Hyp_0:~~X\sim\f_0(X)$\\
\null\hskip0.5cm$\Hyp_1:~~X\sim\f_1(X)$
\vskip0.1cm
\noindent where $\f_0(X),\f_1(X)$ the probability density for $X$ under the nominal and the alternative hypothesis. Both densities are assumed \textit{unknown}. The true densities are as follows: $\f_0(X)$ is a Gaussian vector of length $k=10$, with mean 0 and covariance matrix the identity and $\f_1(X)$ a Gaussian vector of the same length with mean $\frac{1}{\sqrt{10}}\times[1\cdots1]^\intercal$ and covariance matrix the identity times 1.2.
We generate two sets of data, with the same number of samples, $\{X_1^0,\ldots,X_{n}^0\}$ from $\f_0(X)$ and the $\{X_1^1,\ldots,X_{n}^1\}$ from $\f_1(X)$ and use these data to estimate the likelihood ratio, the log-likelihood ratio and posterior probability using the optimization problems defined in the beginning of the current section.

For all three estimators we select the same full 2-layer neural network with configuration $10\times20\times1$. We consider $n=100$ training samples from both densities $\f_0(X),\f_1(X)$. The neural networks are trained using gradient descent as in \eqref{eq:alg1} with simultaneous normalization of the gradient elements following the technique in \cite{tieleman}. We select step-size $\mu=2\times10^{-4}$ and smoothing factor for the normalization $\lambda=0.99$. We run each algorithm for 10,000 iterations.

\begin{figure}[!h]
\centerline{\includegraphics[width=8cm]{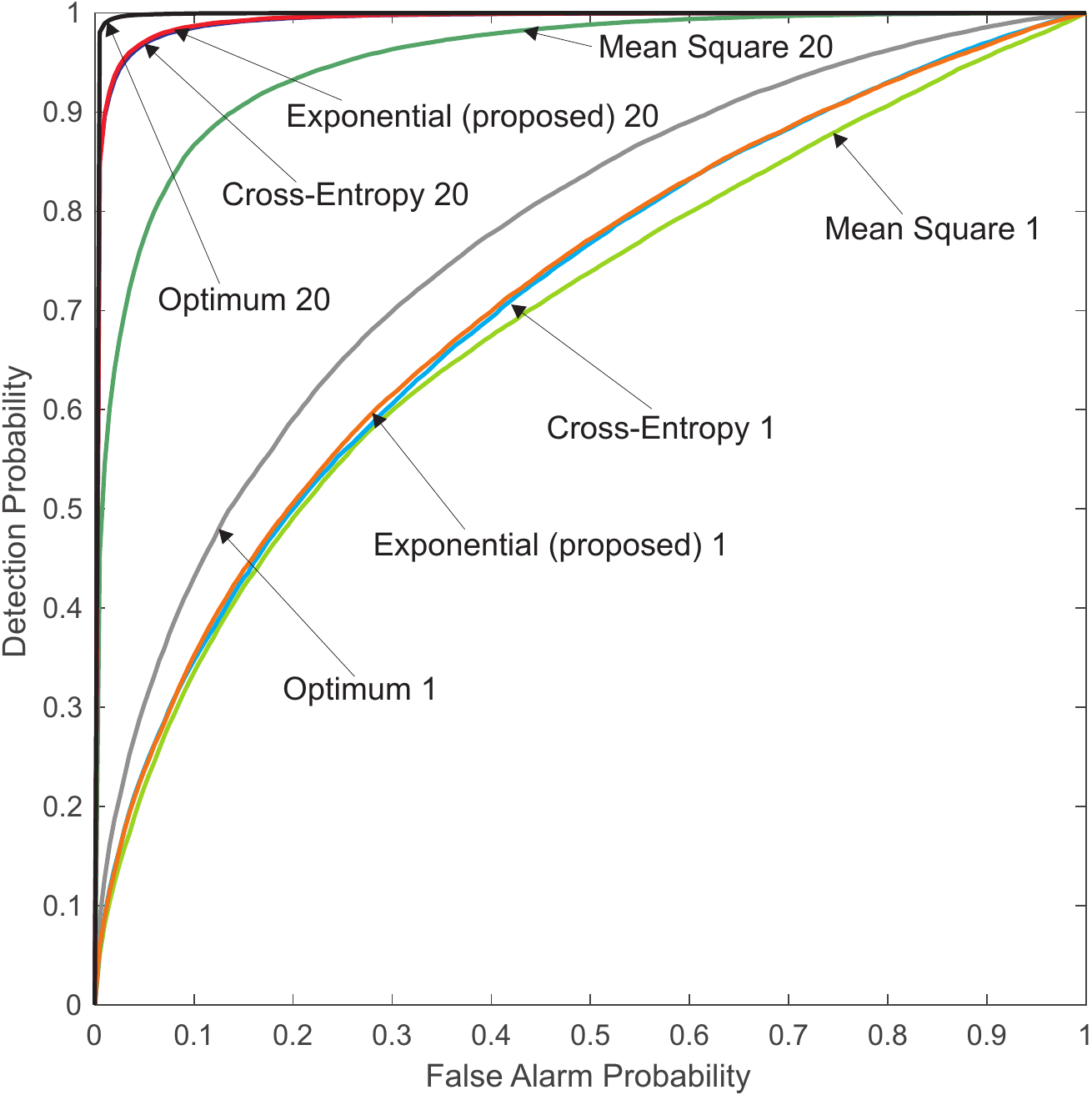}}
\caption{ROC curves for a)~optimum with exact knowledge of densities (gray, black), b)~exponential loss (proposed) for log-likelihood ratio estimation (orange, red), c)~cross-entropy loss for posterior probability estimation (cyan, blue) and d)~mean square loss for likelihood ratio estimation (light green, green).}
\label{fig:2}
\end{figure}
Once the networks are designed we apply them to 100,000 realizations from $\f_0(X)$ and an equal number from $\f_1(X)$ and we use the outcomes to decide between $\Hyp_0$ and $\Hyp_1$ with the help of different thresholds. In Fig.\,\ref{fig:2} we report the corresponding ROC curves. Because we are testing only a single sample, the corresponding curves are marked as ``Mean Square 1'', ``Exponential 1'' and ``Cross-Entropy 1''. We also include the performance of the optimum test with the exact densities which appears as ``Optimum 1''. We can see that all three estimators, at least for interesting values of the false alarm probability, tend to have similar performance. However, these results can be misleading if we use them to evaluate the function estimation quality. If a test appears to have satisfactory performance when tested with \textit{single} samples, this does not necessarily mean that the corresponding neural network estimate is efficient. For example, if we apply a strictly monotone transformation on the exact likelihood ratio and use the outcome for testing we will experience exactly the same performance in both cases. However the transformed version is not necessarily a good ``estimate'' of the original likelihood ratio.

To actually evaluate the quality of the corresponding estimates, instead of using the estimators to test single data samples we use them to test \textit{blocks} of data. Assuming that the samples in each block are i.i.d.~we apply each estimator to every sample in the block, then transform appropriately the resulting values in order to correspond to log-likelihood ratios, which we finally add to form the log-likelihood ratio of the whole block. We select a block size equal to 20 therefore, each experiment requires 20 samples from each of the two densities $\f_0(X)$ and $\f_1(X)$. As before, we repeat the experiment 100,000 times. Fig.\,\ref{fig:2} has the corresponding results with the labels marked with the number ``20'' to indicate the block size. As we can see the classical mean square estimator of the likelihood ratio has clearly inferior performance compared to the other two methods which again are almost indistinguishable.

The reason we did not include the linear and the hinge loss estimators is because they estimate the sign of the log-likelihood ratio function. With the sign it is possible to make decisions, however changing the threshold to produce the ROC curve corresponds to estimating the function $\sgn\big(\log(\sfr-\nu)\big)$ for different thresholds $\nu$. Unfortunately, each $\nu$ requires the determination of its own neural network, which is computationally very demanding.

\subsection{Application to Classification}\label{ssec:4.B}
Let us now apply our method to real data. From the MNIST dataset that contains images of hand-written numerals \cite{lecun}, we isolate the images corresponding to the numbers ``4'' and ``9''. We select this particular pair due to the resemblance of their hand-written versions which makes their distinction challenging. 

We recognize that this is a standard \textit{classification} problem. The difference with detection/hypothesis testing is that we are not interested in the ROC curve but only in minimizing the frequency of classification errors. This corresponds to the Bayesian hypothesis testing for minimum error probability and, for simplicity, we are going to assume that the two prior probabilities of the two classes/hypotheses are equal. We know that the optimum decision mechanism/classifier is the log-likelihood ratio compared to 0 or, equivalently, we can base our decision on the \textit{sign of the log-likelihood ratio}. As we have seen in Section\,\ref{ssec:2.D} we can directly estimate this function using the proposed linear loss and the existing method based on the hinge loss.
Both methods are detailed in the beginning of this section. We also use the mean square estimate of the likelihood ratio and the cross-entropy method that estimates the posterior probability. 

From MNIST we obtain 5500 training images for each of the numerals ``4'' and ``9'', therefore $n_0=n_1=5500$. For testing there are 982 images for ``4'' and 1009 for ``9''. Each image is of size $28\times28$ and we reshape it into a vector of length 784. For the full 2-layer network we select configuration $784\times300\times1$. We apply gradient descent with simultaneous normalization of the gradient elements. The step-size is $\mu=2\times10^{-4}$ and the smoothing factor for gradient normalization is $\lambda=0.99$. We use 1000 iterations.

\begin{figure}[!h]
\centerline{\includegraphics[width=8cm]{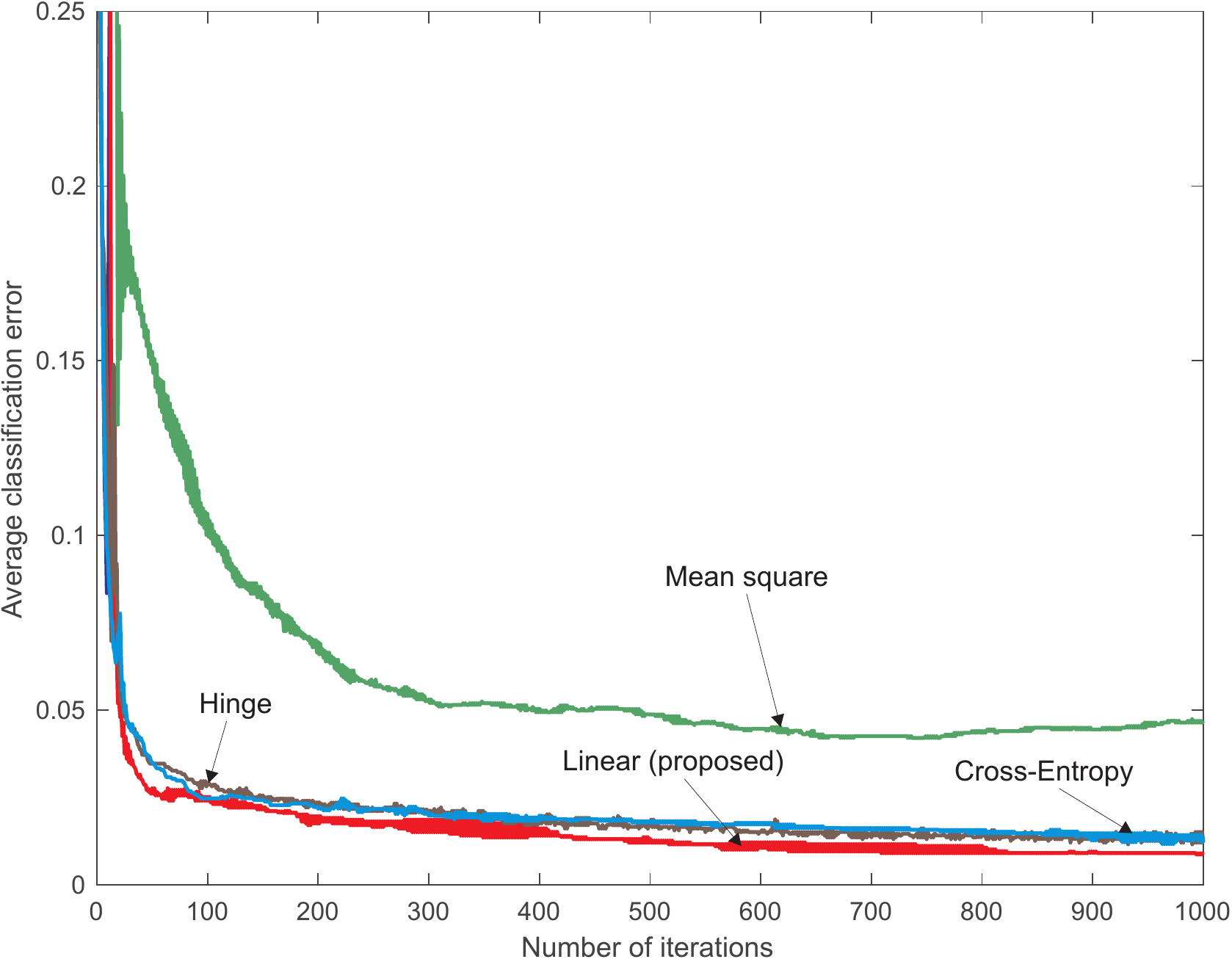}}
\caption{Evolution of the average classification error as a function of the number of iterations of the corresponding training algorithm for: a)~linear loss (proposed) for sign of log-likelihood ratio estimation (red), b)~mean square loss for likelihood ratio estimation (green), c)~hinge loss for sign of log-likelihood ratio estimation (gray) and d)~cross-entropy loss for posterior probability estimation (blue).}
\label{fig:3}
\end{figure}
During training and at \textit{each} iteration, the updated network parameters correspond to a neural network which is applied to the \textit{testing data}. This allows for the computation of the average classification error 
for the \textit{running} version of the neural network. Fig.\,\ref{fig:3} depicts the evolution of this performance index for the four methods. We observe that the proposed linear loss, the hinge loss and the cross-entropy loss enjoy comparable performances. On the other hand decisions based on the estimate of the likelihood ratio, once more, appear to have inferior performance. 

We would like to add that, this particular dataset could also serve as an example for the non-suitability of the parametric density approach to modern data. By consulting the available training images we observe collections of pixels, at fixed locations, that are 0 throughout the whole training set. Unfortunately, determinism in the data can be problematic for methods that rely on randomness. Of course, a simple solution would be to discard these pixels since they bear no information. However, it is also possible the determinism to be hidden. For example, we could have pairs of pixels that are linked to each other through some deterministic formula, while each individual pixel appears as having random content. In the case of a linear relation, a Gaussian modeling would produce an estimate of the covariance matrix which is singular and, therefore, impossible to use for decision making. Proper \textit{pre-processing}, to accommodate parametric approaches, is not straightforward since it is clearly data dependent and not known in advance. On the other hand, neural network methods appear to be robust to this type of problematic data. This is evident from the current simulation where training was applied, blindly, without any regard to zeros or hidden deterministic structures. As we can see the resulting networks successfully distinguish the testing data with small classification error.

\subsection{Data-Driven Equivalent of the Generalized Likelihood Ratio Test}\label{ssec:4.C}
The Generalized Likelihood Ratio Test (GLRT) \cite[Pages 84--88]{moulin} is a possible means to solve the binary hypothesis testing problem when one or both probability densities are under a parametric form. For GLRT we distinguish two versions.

\subsubsection*{\underline{GLRT-1}} We assume that the nominal density $\f_0(\cdot)$ is exactly known and that the alternative $\f_1(\cdot,\vartheta_1)$ is parametrized with $\vartheta_1$. We are given a set of samples $\{X_1,\ldots,X_n\}$ for which we like to decide between the two hypotheses. With GLRT we estimate $\vartheta_1$ with maximum likelihood and replace it in the likelihood ratio. This leads to the statistic
\begin{equation}
\frac{\max_{\vartheta_1}\f_1(X_1,\ldots,X_n,\vartheta_1)}{\f_0(X_1,\ldots,X_n)},
\label{eq:glrt1}
\end{equation}
which we must compare to a threshold in order to make a decision.

\subsubsection*{\underline{GLRT-2}} In this case we assume that both densities are parametrized $\f_0(\cdot,\vartheta_0),\f_1(\cdot,\vartheta_1)$. If we are given a set of samples $\{X_1,\ldots,X_n\}$ to test, we form the likelihood ratio with the two parameter vectors replaced by their maximum likelihood estimates. This means that we compare the statistic
\begin{equation}
\frac{\max_{\vartheta_1}\f_1(X_1,\ldots,X_n,\vartheta_1)}{\max_{\vartheta_0}\f_0(X_1,\ldots,X_n,\vartheta_0)}
\label{eq:glrt2}
\end{equation}
to a threshold. This version clearly becomes problematic when the two densities correspond to the \textit{same} parametric density, namely $\f_0(\cdot,\vartheta_0)=\f(\cdot,\vartheta_0)$ and $\f_1(\cdot,\vartheta_1)=\f(\cdot,\vartheta_1)$. In this case \eqref{eq:glrt2} will produce the same numerator and denominator and the resulting test statistic will be meaningless. A method to resolve this problem is to assume that we have available an additional dataset $\{X_1^0,\ldots,X_{n}^0\}$ sampled from $\f(\cdot,\vartheta_0)$ which can be used to estimate $\vartheta_0$ with, for example, maximum likelihood
$$
\hat{\vartheta}_0=\arg\max_{\vartheta_0}\f(X_1^0,\ldots,X_n^0,\vartheta_0),
$$
and then treat $\f_0(\cdot)$ as known using the estimate $\f_0(\cdot)=\f(\cdot,\hat{\vartheta}_0)$.

Under a data-driven scenario, known densities are replaced with data. Therefore for both versions GLRT-1 and GLRT-2 it makes sense to assume availability of a \textit{fixed} dataset $\{X_1^0,\ldots,X_{n}^0\}$ that represents the nominal density $\f_0(\cdot)$. Every time a new set of samples $\{X_1,\ldots,X_n\}$ is acquired for testing, we are going to \textit{simultaneously estimate and test}, exactly as in GLRT. The difference is that, instead of estimating parameters, we estimate the log-likelihood ratio function.

For simplicity we assume that the samples are i.i.d. but we can also accommodate homogeneous Markov models using the technique proposed in Section\,\ref{ssec:3.B}. Under the i.i.d.~assumption we can write for the log-likelihood ratio
$$
\sfL_n=\log\frac{\f_1(X_1,\ldots,X_n)}{\f_0(X_1,\ldots,X_n)}=\sum_{i=1}^n\log\frac{\f_1(X_i)}{\f_0(X_i)}\approx\hat{\sfL}_n=\sum_{i=1}^n\sfu_1(X_i,\theta_1),
$$
where $\sfu_1(X,\theta_1)$ is a neural network estimate of the log-likelihood ratio of a single sample. This estimate is possible to obtain since there are two datasets available, namely, the fixed set $\{X_1^0,\ldots,X_{n}^0\}$ sampled from the nominal density and $\{X_1,\ldots,X_{n}\}$ which is the set to be tested. 

To evaluate our idea, we consider $X$ to be scalar with $\f_0(X)$ a standard Gaussian and $\f_1(X)$ a Gaussian with mean 0.4 and variance 1.2. We consider two testing scenarios: a)~$n=100$ and b)~$n=200$. For each scenario there exists a fixed collection of $n$ samples representing the nominal density and we generate additional $n$ samples from the nominal and $n$ samples from the alternative in order to test each method with data from both hypotheses. 

\begin{figure}[!h]
\centerline{\includegraphics[width=8cm]{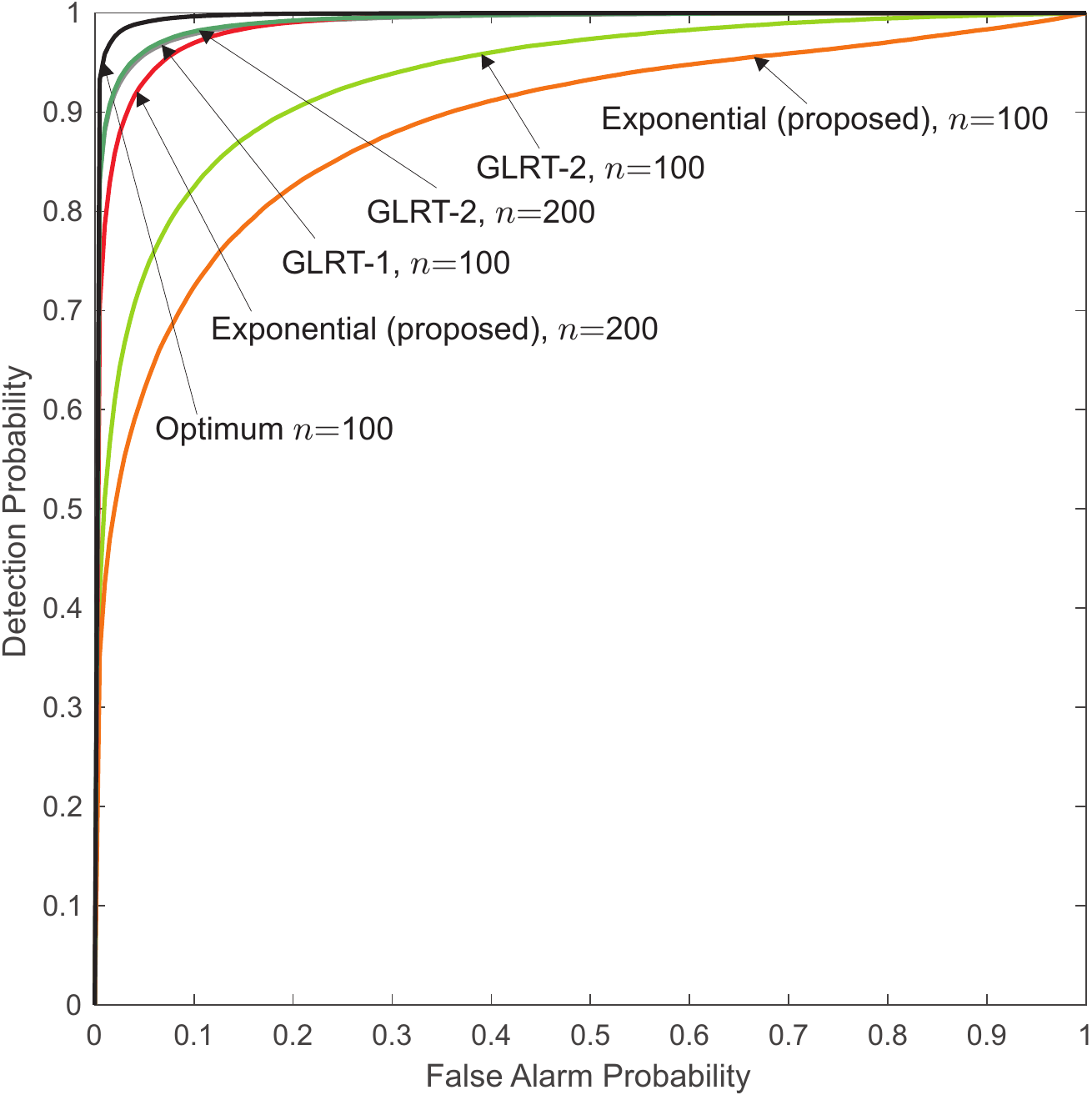}}
\caption{ROC curves for optimum (black) with $n=100$ samples, GLRT-1 (gray) with $n=100$ samples, GLRT-2 with $n=100$ samples (light green) and $n=200$ samples (green) and exponential loss (proposed) neural network based with $n=100$ samples (orange) and $n=200$ samples (red).}
\label{fig:4}
\end{figure}
For the log-likelihood ratio estimation we use the exponential loss method we introduced in the beginning of this section with neural network configuration $1\times20\times1$. We apply gradient descent with step-size $\mu=2\times10^{4}$, smoothing factor $\lambda=0.99$ and 5000 iterations. To find the detection and false alarm probabilities, we repeat the experiment 100,000 times and plot the corresponding ROC curves. Fig.\,\ref{fig:4} depicts our results. In black we have the performance of the optimum test for $n=100$ samples. Here we have exact knowledge of the two densities. Then follows GLRT-1 which knows the nominal density and estimates the mean and variance of the testing data. We plot its performance in gray for $n=100$ samples. The third method is GLRT-2 where we estimate the mean and variance of the nominal from the fixed set and the mean and variance of the testing set. We plot the results for $n=100$ and $n=200$ samples. For $n=100$ we have a notable difference in performance from GLRT-1. However, when we double the data then GLRT-2 with $n=200$ samples matches the performance of GLRT-1 with $n=100$ samples. Finally, we present the results of our method for $n=100$ (orange) and $n=200$ (red) samples. As expected, it is inferior to GLRT-2, but the difference in performance is not dramatic in the case of $n=200$ samples. We should note that GLRT-1 estimates only 2 and GLRT-2 only 4 parameters that parametrize the \textit{exact parabolic form} of the log-likelihood ratio, as opposed to the proposed method which, with the specific configuration, identifies 61 parameters to form a \textit{piecewise linear approximation} of the log-likelihood ratio.

\subsection{Application to Sequential Change Detection}\label{ssec:4.D}
Let us now come to our last simulation. We assume that we are monitoring a random process $\{x_t\}$ which, at some unknown point in time, changes its statistical behavior. The goal is to detect the change as soon as possible using an on-line/sequential test. Sequential change detection finds applications in numerous scientific fields \cite{tartakovsky} with the corresponding literature being vast. We would only like to mention a recent article \cite{khan} that uses deep neural networks to estimate the likelihood ratio in order to perform seizure detection in real EEG signals. In \cite{khan}, acquired data are treated as independent across time allowing for the existing likelihood ratio estimation method to be directly applied.

In our simulation we use synthetic data because, as we mentioned, we would like to observe how the estimation quality of our method affects the overall performance of the sequential detector and how close this performance is to the ideal case of an exactly known data model. For the detection we use the CUSUM test \cite{page} which is based on the CUSUM statistic $\{\sfS_t\}$. The latter is properly approximated by its data-driven version $\{\hat{\sfS}_t\}$ as discussed in Section\,\ref{ssec:3.B}. The CUSUM stopping rule consists in comparing the statistic to a constant threshold $\nu$, therefore, the corresponding CUSUM stopping time and its approximation are defined as
\begin{equation}
\T_{\mathrm{C}}=\inf\{t>0:\sfS_t\geq\nu\},~~~\hat{\T}_{\mathrm{C}}=\inf\{t>0:\hat{\sfS}_t\geq\nu\},
\end{equation} 
where, we recall, that $\sfS_t$ and $\hat{\sfS}_t$ are updated according to \eqref{eq:L12}. $\T_{\mathrm{C}}$ is known to be exactly optimum when the process is i.i.d.~before and after the change \cite{moustakides}, but here we intend to apply the test to Markov data. For Markovian processes before and after the change we know that CUSUM is only asymptotically optimum \cite[Pages 410--417]{tartakovsky}. The data model we adopt is the following: Before the change $\{x_t\}$ is i.i.d.~standard Gaussian, while after the change it becomes conditionally Gaussian of the form $x_t=\mathrm{sign}(x_{t-1})\sqrt{|x_{t-1}|}+w_t$ with $\{w_t\}$ i.i.d.~standard Gaussian. For the log-likelihood ratio estimation \textit{the training data are available before hand} for both pre- and post-change density. As we mentioned, this requirement replaces the classical need for prior knowledge of the pre- and post-change density. The only prior knowledge we have regarding the data model is that the pre- and post-change densities are Markovian of order up to one. Clearly this is not an exact information for the pre-change density since it is i.i.d.

The update of $\hat{\sfS}_t$ requires the estimation of the first and second order log-likelihood ratios $\sfu_1(x_t,\theta_1)$ and $\sfu_2(x_t,x_{t-1},\theta_2)$ defined in \eqref{eq:L11}. The optimization problem is based on the exponential loss defined in the beginning of this section. We use two full 2-layer neural networks with configurations $2\times50\times1$ and $1\times20\times1$. Training is performed with gradient descent with step-size $\mu=2\times10^{-4}$, smoothing factor $\lambda=0.99$ and 10,000 iterations.

\begin{figure}[!h]
\centerline{\includegraphics[width=8cm]{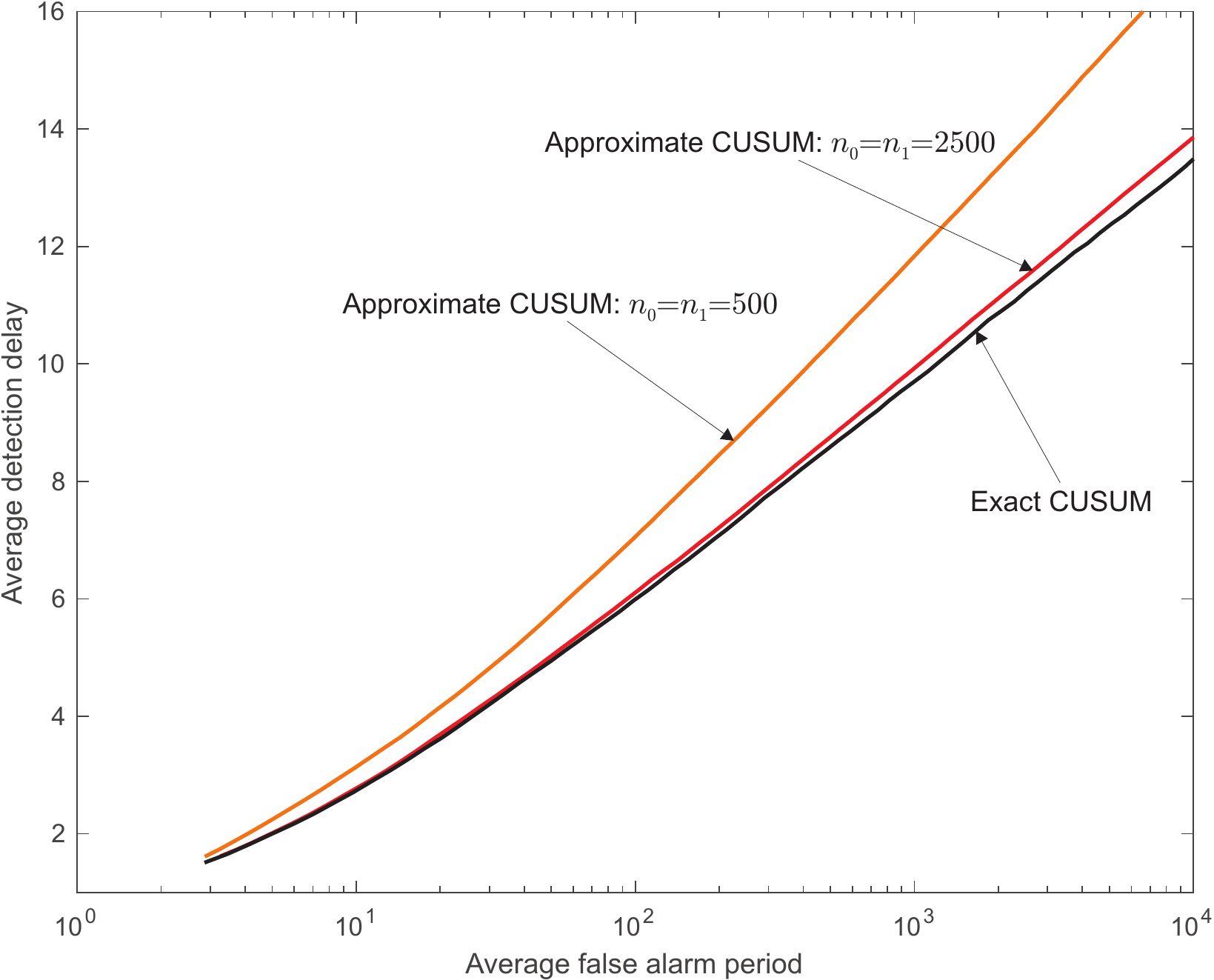}}
\caption{Average detection delay as a function of the average false alarm period, for exact CUSUM (black), approximate CUSUM with the log-likelihood function estimated using proposed with exponential loss with a)~$n_0=n_1=500$ training samples (orange) and b)~with $n_0=n_1=2500$ training samples (red).}
\label{fig:5}
\end{figure}
Once the estimates of the log-likelihood ratios are available we apply the exact CUSUM and the approximate CUSUM test to pre- and post-change data and register the corresponding stopping times $\T_{\rm C}$ and $\hat{\T}_{\rm C}$. The experiment is repeated 100,000 times in order to estimate the average detection delay (for the change occuring at time 0) and the average false alarm period. This procedure is performed for a collection of thresholds $\nu$ in order to generate pairs of average detection delays and average false alarm periods. Fig.\,\ref{fig:5} depicts our simulation results. In black we have the performance of the exact CUSUM where we know exactly the pre- and post-change conditional density. The other two curves correspond to the approximate CUSUM when we estimate the log-likelihood function a)~with $n_0=n_1=500$ training samples (orange) and b) with~$n_0=n_1=2500$ training samples (red). As we can see, with sufficient number of training data, we can approach the exact CUSUM performance very closely.

\begin{remark}
It is somewhat surprising that for exactly the same network configuration, which is rather small sized, we observe such dramatic improvement when going from 500 training samples to 2500. This does not mean that the optimization algorithm failed in any sense, with the 500 samples. The difference is due to the fact that with 2500 samples we have frequent occurrence of ``rare'' events that force the estimator to assign to them efficient estimates of the log-likelihood ratio. In the case of 500 samples when these ``rare'' events are not present the estimator will simply assign arbitrary values. During testing, when the volume of data is significantly larger than the training size (especially with long false alarm periods), ``rare'' events will occur repeatedly and produce wrong updates with the approximate CUSUM statistic. This in turn will result in performance degradation. There is certainly a relationship between average false alarm period and number of training samples that could guarantee some form of asymptotic optimality of the proposed test. But this is, obviously, a theoretical challenge for future research.
\end{remark}

\section{Conclusion}\label{sec:5}
We presented a methodology for designing optimization problems that accept as optimum solution a pre-specified transformation of the likelihood ratio of two probability densities. We then used this method under a data-driven scenario in a variety of well known problems in detection and hypothesis testing. For each of these problems we showed how one can use the optimization problems to obtain a neural network estimate of the likelihood ratio or its transformation when only data sampled from the two densities are available. Simulation results with synthetic and true data indicate that it is a better strategy to estimate the log-likelihood ratio instead of the likelihood ratio which is the current practice in the literature since it results in a noticeable performance improvement. Furthermore, the choice of log-likelihood ratio estimation offers a straightforward solution to the problem of non-negative likelihood ratio estimates which, with the existing techniques, it is a difficult property to satisfy.

\ifCLASSOPTIONcaptionsoff
  \newpage
\fi

\end{document}